\documentclass{jpp}

\usepackage{graphicx}
\usepackage{natbib}
\usepackage{amsmath}

\ifCUPmtlplainloaded \else
  \checkfont{eurm10}
  \iffontfound
    \IfFileExists{upmath.sty}
      {\typeout{^^JFound AMS Euler Roman fonts on the system,
                   using the 'upmath' package.^^J}%
       \usepackage{upmath}}
      {\typeout{^^JFound AMS Euler Roman fonts on the system, but you
                   dont seem to have the}%
       \typeout{'upmath' package installed. JPP.cls can take advantage
                 of these fonts, if you use 'upmath' package.^^J}%
      }
  \else
  \fi
\fi

\ifCUPmtlplainloaded \else
  \checkfont{msam10}
  \iffontfound
    \IfFileExists{amssymb.sty}
      {\typeout{^^JFound AMS Symbol fonts on the system, using the
                'amssymb' package.^^J}%
       \usepackage{amssymb}%
       \let\le=\leqslant  
       \let\ge=\geqslant  
      }{}
  \fi
\fi

\ifCUPmtlplainloaded \else
  \IfFileExists{amsbsy.sty}
    {\typeout{^^JFound the 'amsbsy' package on the system, using it.^^J}%
     \usepackage{amsbsy}}
    {}
\fi

\newcommand\Real{\mbox{Re}} % cf plain TeX's \Re and Reynolds number
\newcommand\Imag{\mbox{Im}} % cf plain TeX's \Im
\newsavebox{\astrutbox}
\sbox{\astrutbox}{\rule[-5pt]{0pt}{20pt}}

\newcommand{\pd}[2]{\ensuremath{\frac{\partial #1}{\partial #2}}}

\newcommand\beq{\begin{equation}}
\newcommand\eeq{\end{equation}}
\newcommand{\bea}{\begin{eqnarray}}
\newcommand{\eea}{\end{eqnarray}}

\newcommand{\figref}[1]{figure~\ref{#1}}

\newcommand{\secref}[1]{section~\ref{#1}}
\newcommand{\secsand}[2]{sections~\ref{#1} and \ref{#2}}

\newcommand{\apref}[1]{appendix~\ref{#1}}

\newcommand{\Eqref}[1]{Equation~(\ref{#1})}
\renewcommand{\eqref}[1]{equation~(\ref{#1})}

\newcommand{\eqsand}[2]{equations~(\ref{#1}) and (\ref{#2})}

\newcommand{\eqsdash}[2]{equations~(\ref{#1}--\ref{#2})}
\newcommand{\exref}[1]{(\ref{#1})}

\newcommand{\lt}{\left}
\newcommand{\rt}{\right}

\newcommand{\la}{\langle}
\newcommand{\ra}{\rangle}
\newcommand{\dd}{\partial}

\renewcommand{\Re}{{\rm Re}}
\newcommand{\sgn}{{\rm sgn}\,}
\newcommand{\eps}{\varepsilon}
\renewcommand{\phi}{\varphi}
\newcommand{\rmd}{{\rm d}}
\newcommand{\vth}{v_{\rm th}}
\newcommand{\geff}{\gamma_{\rm eff}}
\newcommand{\gko}{g_{k\omega}}
\newcommand{\gmko}{g_{m,k\omega}}
\newcommand{\gmk}{g_{m,k}}
\newcommand{\tg}{\tilde g}
\newcommand{\tgmk}{\tilde g_{m,k}}
\newcommand{\tgmko}{\tilde g_{m,k\omega}}
\newcommand{\Cmk}{C_{m,k}}
\newcommand{\phiko}{\phi_{k\omega}}
\newcommand{\chiko}{\chi_{k\omega}}
\newcommand{\chikoone}{\chi_{1,k\omega}}

\title{Fluctuation-dissipation relations for a plasma-kinetic Langevin equation}
\author[A. Kanekar, A. A. Schekochihin, W. Dorland and N. F. Loureiro]%
{A.\ns K\ls A\ls N\ls E\ls K\ls A\ls R,$^1$\thanks{Email address for correspondence: anjor@umd.edu}\ns
A.\ns A.\ns S\ls C\ls H\ls E\ls K\ls O\ls C\ls H\ls I\ls H\ls I\ls N,$^2$\break%
W.\ns D\ls O\ls R\ls L\ls A\ls N\ls D,$^1$ 
\ns \and N.\ns F.\ns L\ls O\ls U\ls R\ls E\ls I\ls R\ls O$^3$}
\affiliation{$^1$Department of Physics, University of Maryland, College Park, MD 20742-3511, USA\\[\affilskip]
$^2$Rudolf Peierls Centre for Theoretical Physics, University of Oxford, Oxford OX1 3NP, UK\\
$^3$Instituto de Plasmas e Fus\~ao Nuclear, Instituto Superior T\'ecnico, Universidade de Lisboa, 
1049-001 Lisbon, Portugal}

\begin{document}

\maketitle
\begin{abstract}
A linearised kinetic equation describing electrostatic perturbations of a Maxwellian equilibrium 
in a weakly collisional plasma forced by a random source is considered. The problem is treated 
as a kinetic analogue of the Langevin equation and the corresponding
fluctuation-dissipation relations are derived. The kinetic fluctuation-dissipation
relation reduces to the standard ``fluid'' one in the regime where the Landau damping rate is small 
and the system has no real frequency; in this case the simplest possible Landau-fluid closure 
of the kinetic equation coincides with the standard Langevin equation.
Phase mixing of density fluctuations and emergence of fine scales in velocity space 
is diagnosed as a constant flux of free energy in Hermite space; 
the fluctuation-dissipation relations for the perturbations of the distribution 
function are derived, in the form of a universal expression for the Hermite spectrum of 
the free energy. Finite-collisionality effects are included. 
This work is aimed at establishing the simplest fluctuation-dissipation relations for 
a kinetic plasma, clarifying the connection between Landau and Hermite-space formalisms, 
and setting a benchmark case for a study of phase mixing in turbulent plasmas.
\end{abstract}

\section{Introduction}
Fluctuation dissipation relations (FDR) predict the response of a dynamical system to an externally 
applied perturbation, based on the system's internal dissipation properties. The classical Langevin
equation \citep{kubo66} supplies the best known example of such FDR. 
The standard formulation is to consider a scalar $\phi$ 
forced by a Gaussian white-noise source $\chi$ and damped at the rate $\gamma$: 
\begin{align}
&\frac{\dd\phi}{\dd t} + \gamma \phi  = \chi, \label{eq:Langevin} \\
&\la \chi(t) \chi(t') \ra  = \eps \delta (t-t'), \nonumber
\end{align}
where angle brackets denote the ensemble average and 
$\eps/2$ is the mean power injected into the system by the source:
\beq
\frac{\rmd}{\rmd t}\frac{\la\phi^2\ra}{2} + \gamma\la\phi^2\ra = \frac{\eps}{2}.  
\eeq
The steady-state mean square fluctuation level is then given by the FDR, linking 
the injection and the dissipation of the scalar fluctuations:    
\beq
\la \phi^2 \ra = \frac{\eps}{2\gamma}. 
\label{eq:FDR}
\eeq

The simplest physical example of such a system is a Brownian particle suspended 
in liquid, with $\phi$ the velocity of the particle and $\gamma$ the frictional damping.
More generally, \eqref{eq:Langevin} may be viewed as a generic model for 
systems where some perturbed quantity is randomly stirred and decays via 
some form of linear damping, a frequently encountered situation in, e.g., 
fluid dynamics. 

Nearly every problem in plasma physics involves a system with 
driven and damped linear modes. Here we consider the prototypical 
such case: the behaviour of perturbations of a Maxwellian equilibrium 
in a weakly collisional plasma in one spatial and one velocity-space dimension.
In such a system (and in weakly collisional or collisionless plasmas generally), 
damping of the perturbed electric fields occurs via the famous \citet{landau46} 
mechanism. Landau damping, however, is different in several respects from 
standard ``fluid'' damping phenomena. It is in fact a phase mixing process: 
electric---and, therefore, density---perturbations are phase mixed 
and thus are effectively damped. Their (free) energy is transferred 
to perturbations of the particle distribution function that develop 
ever finer structure in velocity space and are eventually removed by 
collisions or, in a formally collisionless limit, by some suitable coarse-graining procedure.  
The electrostatic potential $\phi$ in such systems cannot in general be 
rigorously shown to satisfy a ``fluid'' equation of the form \exref{eq:Langevin}, 
with $\gamma$ the Landau damping rate, although the idea that \eqref{eq:Langevin} 
or a higher-order generalisation thereof 
is not a bad model underlies the so-called Landau-fluid closures 
\citep{hammett90,hammett92,hedrick92,dorland93,snyder97,passot04,goswami05,passot07}. 

It is a natural question to ask whether, despite the dynamical equations
for $\phi$ (or, more generally, for the moments of the distribution function) 
being more complicated than \eqref{eq:Langevin}, we should still expect the 
mean fluctuation level to satisfy \eqref{eq:FDR}, where $\gamma$ is 
the Landau damping rate. And if that is not the case, then should the value of 
$\gamma$ {\em defined} by \eqref{eq:FDR} be viewed as the 
effective damping rate in a driven system, replacing the Landau rate? 
\citet{plunk13} recently considered the latter question
and argued that the fact that the effective damping rate defined this way 
differs from the Landau rate suggests a fundamental modification of Landau response 
in a stochastic setting. Our take on the problem at hand differs 
from his somewhat in that we take the kinetic version of the Langevin 
equation (introduced in \secref{sec:kin}) at face value and 
derive the appropriate kinetic generalisation of the FDR, 
instead of attaching a universal physical significance to the ``fluid'' version of it. 
Interestingly, the kinetic FDR does simplify to the classical fluid FDR when 
the Landau damping rate is small. Furthermore, we prove that in this limit 
(and when the system has no real frequency), the dynamics of $\phi$ is in fact 
described by \eqref{eq:Langevin} 
with $\gamma$ equal precisely to the Landau rate (i.e., the simplest Landau fluid 
closure is a rigorous approximation in this limit). 
The latter result is obtained by treating the velocity-space dynamics 
of the system in Hermite space. We also show how phase mixing in our system can 
be treated as a free-energy flux in Hermite space, what form the FDR takes 
for the Hermite spectrum of the perturbations of the distribution function, 
and how collisional effects can be included. The intent of this treatment 
is to provide a degree of clarity as to the behaviour of a very simple 
plasma model and thus set the stage for modelling more complex, nonlinear 
phenomena. 

The plan of the paper is as follows. In \secref{sec:kin}, 
we describe a simple model for a weakly collisional plasma, which we call the
kinetic Langevin equation, and then, in \secref{sec:FDR}, derive the FDR for the same, 
including the ``fluid'' limit mentioned above. In \secref{sec:Hermite}, 
Hermite-space dynamics are treated, including the limit where 
Landau-fluid closures hold rigorously.  
An itemised summary of our findings is given in \secref{sec:disc}. A version of our
calculation with a different random source is presented in \apref{ap:g1force}.

\section{Kinetic Langevin equation}
\label{sec:kin}

We consider the following (1+1)-dimensional model of a homogeneous plasma 
perturbed about a Maxwellian equilibrium: 
\begin{align}
&\pd{g}{t} + \underbrace{v\,\pd{g}{z}}_{\text{phase mixing}} 
+ \underbrace{v F_0 \pd{\phi}{z}}_{\text{electric field}} = 
\underbrace{\chi(t) F_0}_{\text{source}} +
\underbrace{C[g]}_{\text{collisions}}\!\!\!, \label{eq:g} \\
&\phi  = \alpha \int_{-\infty}^\infty \rmd v\, g, \label{eq:phi} \\
& \la \chi(t) \chi(t') \ra  = \eps \delta (t-t'), \nonumber
\end{align}
where $g(z,v,t)$ is the perturbed distribution 
function and $F_0(v)$ is the Maxwellian equilibrium distribution
$F_0=e^{-v^2}/\sqrt{\pi}$. The velocity $v$ (in the $z$ direction) 
is normalised to the thermal speed $\vth = \sqrt{2T/m}$ 
($T$ and $m$ are the temperature and mass of the particle species under consideration), 
spatial coordinate $z$ is normalised to an arbitrary length $L$, and time $t$ to $L/\vth$.
Only one species (either electrons or ions) is evolved. 
The second species follows the density fluctuations of the first via 
whatever response a particular physical situation warrants: 
Boltzmann, isothermal, or no response---all of these possibilities 
are embraced by \eqref{eq:phi}, which determines the (suitably normalised) 
scalar potential $\phi$ in terms of the perturbed density associated with~$g$; 
the parameter $\alpha$ contains all of the specific physics. 
For example, if $g$ is taken to be the perturbed ion distribution function 
in a low-beta magnetised plasma and electrons to have Boltzmann response, 
then $\alpha=ZT_e/T_i$, the ratio of the electron to ion temperatures ($Z$ is the ion charge in 
units of electron charge~$e$)---the resulting system describes (Landau-damped) ion-acoustic waves;
\eqref{eq:phi} in this case is the statement of quasineutrality. 
Another, even more textbook example is damped Langmuir waves, the case 
originally considered by \citet{landau46}: $g$ is the perturbed electron distribution function, 
ions have no response, so $\alpha=2/k^2\lambda_{D}^2$, where 
$\lambda_{D}$ is the Debye length and $k$ is the wave number of the 
perturbation ($\dd/\dd z = ik$); \eqref{eq:phi} in this case is the Gauss-Poisson law. 

A particularly astrophysically and space-physically relevant example 
(in the sense of being accessible to measurements in the solar wind; e.g., 
\citealt{celnikier83,celnikier87,marsch90,bershadskii04,hnat05,chen11}) is 
the compressive perturbations in a magnetised plasma---perturbations 
of plasma density and magnetic-field strength at scales long compared 
to the ion Larmor radius. These are in fact described 
by two equations evolving two decoupled 
functions $g^+$ and $g^-$, which are certain linear combinations 
of the zeroth and second moments of the perturbed ion distribution 
function with respect to the velocity perpendicular to the 
mean magnetic field (taken to be in the $z$ direction). 
These equations are derived in \citet[][\S 6.2.1]{tome} 
and are of the form \exref{eq:g} with 
\beq
\alpha^\pm =-\lt[-\frac{T_i}{ZT_e} + \frac{1}{\beta_i}\pm A\rt]^{-1}, \quad
A = \sqrt{\lt(1+\frac{T_i}{ZT_e}\rt)^2 + \frac{1}{\beta_i^2}} 
\eeq
for $g^\pm$, respectively (here $\beta_i=8\pi n_iT_i/B^2$ is the ion beta). 
The physical fields, the density and magnetic-field-strength perturbations, are related 
to $g^\pm$ by 
\begin{align}
\frac{\delta n}{n} &= \frac{1}{2A}\int\rmd v\lt[\lt(1 + \frac{T_i}{ZT_e} + \frac{1}{\beta_i} + A\rt)g^-
- \frac{T_i}{ZT_e}\frac{2}{\beta_i}\,g^+\rt],\\
\frac{\delta B}{B} &= \frac{1}{2A}\int\rmd v\lt[\lt(1 + \frac{T_i}{ZT_e} + \frac{1}{\beta_i} + A\rt)g^+
- \lt(1+\frac{ZT_e}{T_i}\rt)g^-\rt].
\end{align}
While these expressions are perhaps not very physically transparent, it may aid 
intuition to note that ${\delta n}/{n} \approx \int\rmd v g^-$ and 
${\delta B}/{B} \approx \int\rmd v g^+$ either in the limit of 
high $\beta_i$ and hot ions ($T_i\gg T_e$) or in the limit of 
low $\beta_i$ and cold ions ($T_i\ll T_e$). At low $\beta_i$, the $g^-$ equation 
describes ion-acoustic waves ($\alpha^-\approx ZT_e/T_i$; see above). 
At high $\beta_i$, the $g^+$ equation describes a kinetic version of the MHD slow mode, 
subject to a version of Landau damping due to \citet{barnes66}; 
in this case, $\alpha^+\approx -1 + 1/\beta_i$. 

Thus, \eqsand{eq:g}{eq:phi} correspond to a variety of interesting physical situations. 

The energy injection in \eqref{eq:g} is modelled by a white-in-time, Maxwellian-in-velocity-space 
source $\chi(t) F_0$ supplying fixed power $\propto \eps$ to the perturbations (see below). 
This is a direct analogue of the noise term in the ``fluid'' Langevin equation \exref{eq:Langevin}
and so this particular choice of forcing was made in order to enable the simplest possible
comparison with the ``fluid'' case.\footnote{One might argue that this is not, however, 
the most physical form of forcing and that it would be better to inject energy 
by applying a random electric field to the plasma, rather than a source of density 
perturbations. In \apref{ap:g1force} we present a version of our calculation for 
such a more physical source, and show that all the key results are similar. Note that 
the forcing in \eqref{eq:g} does not violate particle conservation because 
we assume that spatial integrals of all perturbations vanish: 
$\int\rmd z\, g = 0$, $\int\rmd z\,\chi = 0$.}
The energy injection leads to sharp gradients in the velocity space (phase mixing), 
which are removed by the collision operator $C[g]$. 
``The energy'' in the context of a kinetic equation is the free energy of the 
perturbations \citep[see][and references therein]{schekochihin08,tome}, given in this case by 
\beq
W = \int\rmd v\,\frac{\la g^2\ra}{2F_0} + \frac{\la\phi^2\ra}{2\alpha}
\label{eq:W}
\eeq    
and satisfying 
\beq
\frac{\rmd W}{\rmd t} = \frac{1+\alpha}{2}\,\eps + \int\rmd v\,\frac{\la g C[g]\ra}{F_0}.
\label{eq:Wbalance}
\eeq
The first term on the right-hand side is the energy injection by the source, 
the second, negative definite, term is its thermalisation by collisions.
Note that the variance of $\phi$ is not by itself a conserved quantity:
\beq
\frac{\rmd}{\rmd t}\frac{\la\phi^2\ra}{2} + \alpha\lt\la\phi\frac{\dd}{\dd z}\int\rmd v\,vg\rt\ra 
= \frac{\alpha^2}{2}\,\eps.
\label{eq:phisq}
\eeq
The power $\alpha^2\eps/2$ injected into fluctuations of $\phi$ is 
transferred into higher moments of $g$ via phase mixing. 
Landau damping is precisely this process of draining free energy 
from the lower moments and transferring it into higher moments of the 
distribution function---without collisions, this is just a redistribution 
of free energy within \eqref{eq:W}, which, in the absence of source, 
would look like a linear damping of $\phi$.\footnote{Note that 
$\alpha=-1$ corresponds to an effectively undriven system; 
the Landau damping rate for this case is zero (\eqref{eq:omega_small}).
We will see in \secref{sec:KE_Hermite} that in this case the 
driven density moment decouples from the rest of the perturbed 
distribution function; see \eqref{eq:g1}. For $\alpha<-1$, the system 
is no longer a driven-damped system; this parameter regime never occurs 
physically.} 
 
In the presence of a source, the system described by \eqsand{eq:g}{eq:phi} 
is a driven-damped system much like the Langevin equation \exref{eq:Langevin}. 
The damping of $\phi$ in the kinetic case is provided by Landau damping 
(phase mixing) as opposed to the explicit dissipation term in \eqref{eq:Langevin}. 
It is an interesting question whether 
in the steady state, the second term on the left-hand side 
of \eqref{eq:phisq} can be expressed as $\geff\la\phi^2\ra$, leading 
an analogue of the FDR (\eqref{eq:FDR}), and if so, whether the 
``effective damping rate'' $\geff$ in this expression is equal 
to the Landau damping rate $\gamma_L$. 
The answer is that an analogue of the FDR does exist, 
$\geff$ is non-zero for vanishing collisionality, but in general, $\geff\neq\gamma_L$. 

\begin{figure}
\begin{center}
\includegraphics[width=10cm]{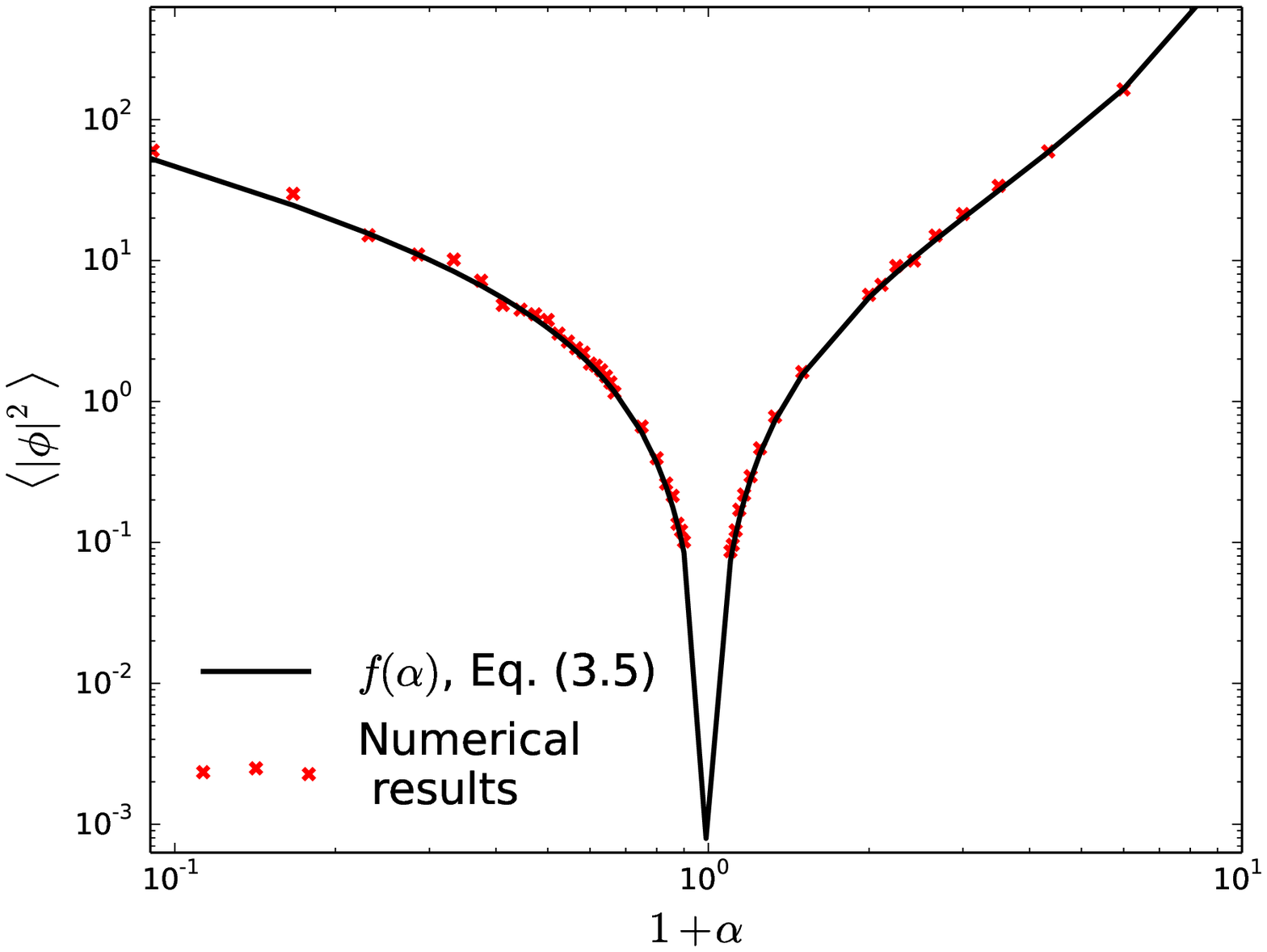}
\caption{Normalised steady-state amplitude $2\pi|k|\la|\phi_k|^2\ra/\eps_k=f(\alpha)$ 
vs.\ $1+\alpha$: the solid line is the analytical prediction 
($f(\alpha)$ as per \eqref{eq:f}), the crosses are computed from 
the long-time limit of $\la|\phi_k|^2\ra$ obtained via direct numerical 
solution of \eqsand{eq:g}{eq:phi}.}
\label{fig:f}
\end{center}
\end{figure}

\section{Kinetic FDR}
\label{sec:FDR}

Ignoring collisions in \eqref{eq:g} and Fourier-transforming it in space and time, 
we~get 
\beq
\gko = - \phiko\frac{v F_0}{v - \omega/k} - \frac{i\chiko}{k}\frac{F_0}{v - \omega/k}.
\label{eq:gko}
\eeq
Introducing the plasma dispersion function $Z(\zeta) = \int\rmd v F_0/(v-\zeta)$, 
where the integration is along the Landau contour \citep{fried61}, we find from 
\eqsand{eq:gko}{eq:phi}:
\begin{align}
\label{eq:phiko}
&\phiko = - \frac{i\chiko}{|k|}\frac{Z(\omega/|k|)}{D_\alpha(\omega/|k|)},\\
&D_\alpha\!\lt(\frac{\omega}{|k|}\rt) = 1 + \frac{1}{\alpha} + \frac{\omega}{|k|} Z\!\lt(\frac{\omega}{|k|}\rt).
\end{align}
Note that $D_\alpha(\omega/|k|) = 0$ is the dispersion relation for the classic \citet{landau46} problem. 
We now inverse Fourier transform \eqref{eq:phiko} back into the time domain, 
\beq
\phi_k(t) = \int\rmd\omega\, e^{-i\omega t}\phiko = 
 - \frac{i}{|k|}\int\rmd\omega\, e^{-i\omega t}\chiko\frac{Z(\omega/|k|)}{D_\alpha(\omega/|k|)},
\eeq
and compute $\la|\phi_k|^2\ra$ 
in the steady state. In order to do this, we use the fact that 
$\chiko \equiv \int\rmd t e^{i\omega t}\chi_k(t)/2\pi$ satisfies 
$\la\chiko\chi^*_{k\omega'}\ra = \eps_k\delta(\omega-\omega')/2\pi$
because $\la\chi_k(t)\chi_k^*(t')\ra = \eps_k\delta(t-t')$, where 
$\eps_k$ is the source power at wave number $k$. 
The result is
\beq
\la|\phi_k|^2\ra = \frac{\eps_k}{2\pi|k|}f(\alpha),\quad 
f(\alpha) = \int_{-\infty}^{+\infty}\rmd\zeta\lt|\frac{Z(\zeta)}{D_\alpha(\zeta)}\rt|^2,
\label{eq:f}
\eeq
where we have changed the integration variable to $\zeta=\omega/|k|$. 
This is the fluctuation-dissipation relation for our kinetic system that predicts the
long-time behaviour of the electrostatic potential. 
The function $f(\alpha)$, computed numerically as per \eqref{eq:f}, 
is plotted in \figref{fig:f}, together with the results of 
direct numerical solution of \eqsand{eq:g}{eq:phi}, in which $f(\alpha)$ 
is found by computing the saturated fluctuation level $\la|\phi_k|^2\ra$. 

\begin{figure}
\begin{center}
\includegraphics[width=10cm]{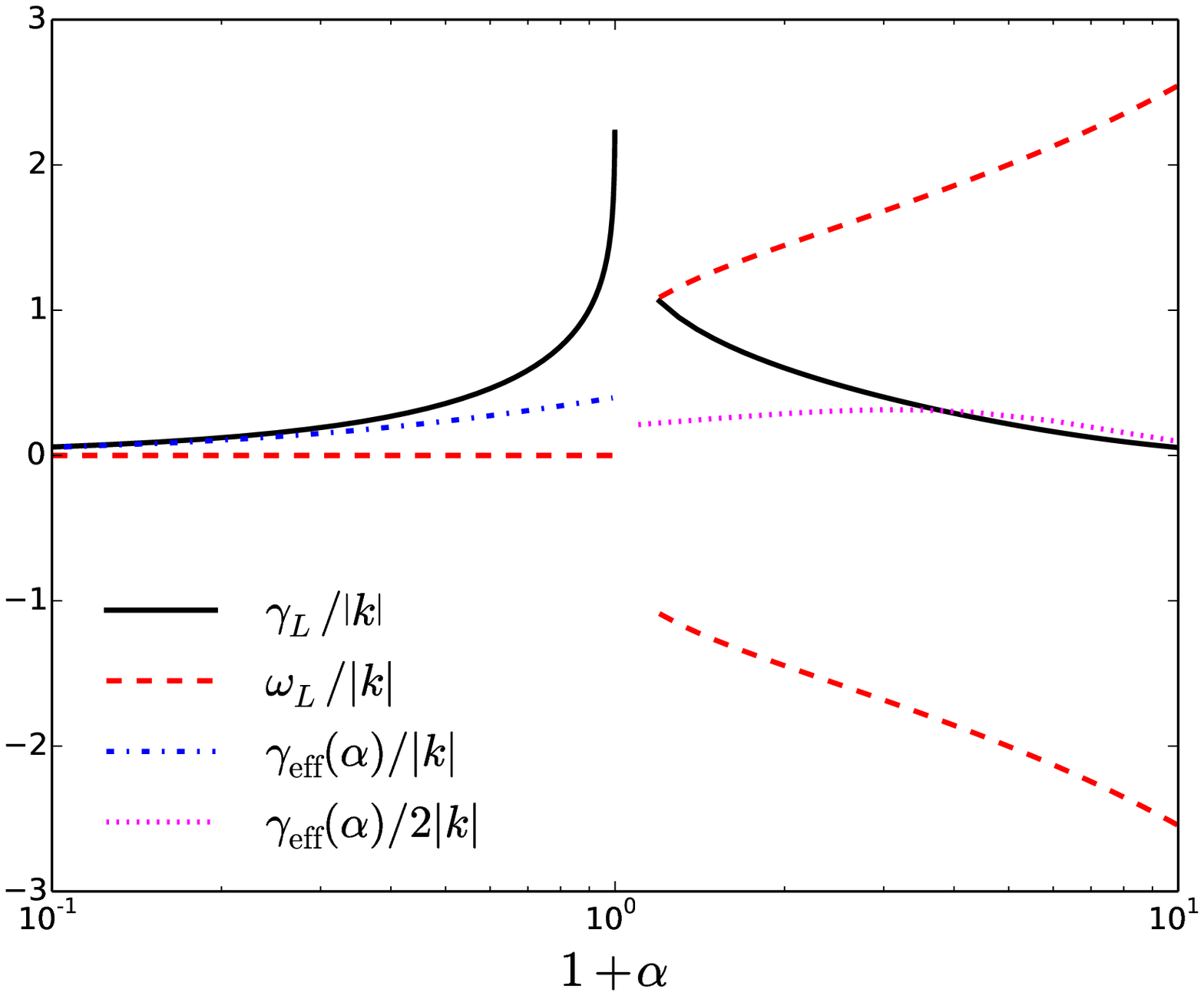}
\caption{Slowest-damped solutions of the dispersion relation 
$D_\alpha(\omega/|k|)=0$: normalised frequency $\omega_L/|k|$ 
(red dashed line) and damping rate $\gamma_L/|k|$ (black soloid line)
vs.\ $1+\alpha$. Also shown are $\geff(\alpha)$ for $\alpha<0$ (blue dash-dotted 
line) and $\geff(\alpha)/2$ for $\alpha>0$ (magenta dotted line), as per 
\eqref{eq:gamma}. The two asymptotic limits in which these 
match $\gamma_L$ are discussed in \secsand{sec:small}{sec:large}.}
\label{fig:gamma_omega}
\end{center}
\end{figure}

\Eqref{eq:f} can be written in the form 
\beq
\la|\phi_k|^2\ra = \frac{\alpha^2\eps_k}{2\geff},\quad 
\geff(\alpha) = \frac{\pi\alpha^2}{f(\alpha)}|k|,
\label{eq:gamma}
\eeq  
but the ``effective damping rate'' $\geff$ is not in general the same 
as the Landau damping rate $\gamma_L$.
This is illustrated in \figref{fig:gamma_omega}, where we plot 
the real ($\omega_L$) and imaginary ($-\gamma_L$) parts of the slowest-damped root(s) 
of $D_\alpha(\omega/|k|) = 0$ 
together with $\geff(\alpha)$ for $\alpha<0$ and $\geff(\alpha)/2$ for $\alpha>0$. 
In the latter case, the linear modes of the system have real frequencies and 
the analogy with the Langevin equation \exref{eq:Langevin} is not apt---a better 
mechanical analogy is a damped oscillator, as explained at the end of 
\secref{sec:large}; the FDR in this case acquires an extra factor of $1/2$, which is 
why we plot $\geff/2$ (see \eqref{eq:FDR_osc}). Remarkably, $\geff(\alpha)$ does 
asymptote to $\gamma_L$ in the limit $1+\alpha\ll1$ and to $2\gamma_L$ in the limit 
$\alpha\to\infty$, i.e., when the damping is weak. 
These asymptotic results can be verified analytically. 

\subsection{Zero real frequency, weak damping ($\alpha\to-1$)} 
\label{sec:small}

When $\alpha+1\ll1$, the solution of the dispersion relation will satisfy 
$\zeta=\omega/|k|\ll1$. In this limit, 
\beq
Z(\zeta)\approx i\sqrt{\pi},\quad
D_\alpha(\zeta)\approx 1 + \frac{1}{\alpha} + i\zeta\sqrt{\pi} 
\approx i\sqrt{\pi}\lt(\zeta + i\frac{1+\alpha}{\sqrt{\pi}}\rt).
\label{eq:Da_small}
\eeq
Therefore, the solution of $D_\alpha(\omega/|k|)=0$ is 
\beq
\omega \approx -i\gamma_L, \quad
\gamma_L = \frac{1+\alpha}{\sqrt{\pi}}|k|. 
\label{eq:omega_small}
\eeq
A useful physical example of Landau damping in this regime is the \citet{barnes66} damping 
of compressive fluctuations in high-beta plasmas, where $1+\alpha\approx 1/\beta_i$ 
\citep[][their equation~(190); see discussion in our \secref{sec:kin}]{tome}. 

Since the zeros of $D_\alpha(\zeta)$ and $D_\alpha^*(\zeta)$, which are 
poles of the integrand in the expression for $f(\alpha)$ (\eqref{eq:f}), 
lie very close to the real line in this case, the integral is easily 
computed by using the approximate expressions \exref{eq:Da_small} for $Z(\zeta)$
and $D_\alpha(\zeta)$ and applying Plemelj's formula, to obtain 
\beq
f(\alpha) \approx \frac{\pi\sqrt{\pi}}{1+\alpha} = \frac{\pi|k|}{\gamma_L}
\quad\Rightarrow\quad
\la|\phi_k|^2\ra\approx \frac{\sqrt{\pi}\eps_k}{2(1+\alpha)|k|} = \frac{\eps_k}{2\gamma_L}.
\label{eq:FDR_small}
\eeq
Since $\alpha^2\approx1$, this is the same as 
\eqref{eq:gamma} with $\geff=\gamma_L$, so the ``fluid'' FDR is recovered.  
Note, however, that this recovery of the exact form of the ``fluid'' FDR 
is a property that is not universal with respect to the exact form of 
energy injection: as shown in \apref{ap:g1force}, it breaks down for 
a different forcing (see \eqref{eq:FDR_small1}).

\subsection{Large real frequency, weak damping ($\alpha\to\infty$)} 
\label{sec:large}
 
Another analytically tractable limit is $\alpha\gg1$, in which case the solutions of the 
dispersion relation have $\zeta=\omega/|k|\gg1$. In this limit, 
\beq
Z(\zeta)\approx i\sqrt{\pi}\,e^{-\zeta^2} - \frac{1}{\zeta} - \frac{1}{2\zeta^3},\quad
D_\alpha(\zeta)\approx \frac{1}{\alpha} - \frac{1}{2\zeta^2} + i\sqrt{\pi}\,\zeta e^{-\zeta^2}.
\label{eq:Da_large}
\eeq
The solutions of $D_\alpha(\omega/|k|)=0$ are 
\beq
\omega \approx \pm\sqrt{\frac{\alpha}{2}}\,|k| - i\gamma_L,\quad
\gamma_L = \sqrt{\pi}\,\frac{\alpha^2}{4}\,e^{-\alpha/2}|k|.
\label{eq:omega_large}
\eeq
Two textbook examples of Landau-damped waves in this regime are 
ion acoustic waves at $\beta_i\ll1$, $T_i\ll T_e$ (cold ions), for which $\alpha=ZT_e/T_i$, 
and long-wavelength Langmuir waves, for which $\alpha = 2/k^2\lambda_D^2$ \citep{landau46}. 

In the integral in \eqref{eq:f}, the poles are again very close to the real line 
and so in the integrand, we may approximate, in the vicinity of one of the 
two solutions \exref{eq:omega_large} 
\beq
Z(\zeta) \approx \mp \sqrt{\frac{2}{\alpha}},\quad
D_\alpha(\zeta) \approx \pm \lt(\frac{2}{\alpha}\rt)^{3/2}\lt(\zeta \mp \sqrt{\frac{\alpha}{2}} + i\,\frac{\gamma_L}{|k|}\rt). 
\eeq
Using again Plemelj's formula and noting that equal contributions arise from each 
of the two roots, we find 
\beq
f(\alpha) \approx 2\sqrt{\pi}\,e^{\alpha/2} 
= \frac{\pi\alpha^2|k|}{2\gamma_L}
\quad\Rightarrow\quad
\la|\phi_k|^2\ra \approx \frac{\alpha^2\eps_k}{4\gamma_L}, 
\eeq
which is the same as \eqref{eq:gamma} with $\geff = 2\gamma_L$. 

Despite the apparently discordant factor of 2, this, in fact, is again 
consistent with a non-kinetic, textbook FDR. However, since we are considering a system with 
a large frequency, the relevant mechanical analogy is not \eqref{eq:Langevin}, but 
the equally standard (and more general) equation for a forced and damped oscillator: 
\beq
\ddot\phi + \gamma\dot\phi + \omega^2\phi = \dot\chi,
\label{eq:osc}
\eeq 
where overdots mean time derivatives. 
At the risk of outraging a mathematically fastidious reader, 
we continue to consider $\chi$ a Gaussian white noise satisfying 
$\la\chi(t)\chi(t')\ra = \eps\delta(t-t')$.
For $\omega=0$, \eqref{eq:osc} then precisely reduces to \eqref{eq:Langevin}. 
For $\omega\neq0$, it is not hard to show (by Fourier transforming in time, 
solving, then inverse Fourier transforming and squaring the amplitude) that 
the stationary mean square amplitude $\la\phi^2\ra$ for \eqref{eq:osc} 
still satisfies \eqref{eq:FDR}. 
However, the relationship between the actual linear damping rate $\gamma_L$ 
of $\phi$ and the parameter $\gamma$ depends on the frequency: 
$\gamma_L = \gamma$ when $\omega\ll\gamma$ and $\gamma_L=\gamma/2$ when $\omega\ge\gamma/2$.  
In the latter case, which is the one with which we are preoccupied here, 
\eqref{eq:FDR} becomes, in terms of $\gamma_L$:  
\beq
\la\phi^2\ra = \frac{\eps}{4\gamma_L}.
\label{eq:FDR_osc}
\eeq
The required extra factor of 2 is manifest.\footnote{As in \secref{sec:small}, 
this very simple mechanical analogy also breaks down for a different choice of forcing; 
see \apref{ap:g1force} (\eqref{eq:FDR_large1}).}

\section{Velocity-space structure}
\label{sec:Hermite}

The kinetic FDR derived in the previous section was concerned 
with the rate of removal of free energy from the density moment 
of the perturbed distribution function. This free energy flows into 
higher moments, i.e., is ``phase mixed'' away. In this section, 
we diagnose the velocity-space structure of the fluctuations
and extend the FDR to compute their amplitude. 

\subsection{Kinetic equation in Hermite space}
\label{sec:KE_Hermite}

The emergence of ever finer velocity-space scales is made explicit
by recasting the kinetic equation \exref{eq:g} in Hermite space, 
a popular approach for many years 
\citep{armstrong67,grant67a,hammett93,parker95,ng99,watanabe04,zocco11,loureiro13,hatch13,plunk14}. 
The distribution is decomposed into Hermite moments as follows
\beq
g(v) = \sum_{m=0}^\infty \frac{H_m (v) F_0 }{\sqrt{2^m m!}} g_m, \quad
g_m = \int \rmd v\, \frac{H_m(v)}{\sqrt{2^m m!}}\, g(v), 
\eeq
where $H_m(v)$ is the Hermite polynomial of order $m$. 
In terms of Hermite moments, \eqref{eq:phi} becomes
\beq
\phi = \alpha g_0,
\label{eq:phi_g0}
\eeq
while \eqref{eq:g} turns into a set of equations for the Hermite moments $g_m$, 
where phase mixing is manifested by the coupling of higher-$m$ moments 
to the lower-$m$ ones: 
\begin{align} 
\label{eq:g0}
&\pd{g_0}{t} + \pd{}{z}\frac{g_1}{\sqrt{2}}  = \chi,\\
\label{eq:g1}
&\pd{g_1}{t} + \pd{}{z}\lt(g_2 + \frac{1+\alpha}{\sqrt{2}}\,g_0\rt)  = 0,\\
&\pd{g_m}{t} + \pd{}{z}\lt(\sqrt{\frac{m+1}{2}}\,g_{m+1} + \sqrt{\frac{m}{2}}\,g_{m-1}\rt) 
= -\nu m g_m,  \quad m\ge2,
\label{eq:gmeq}
\end{align}
where $\nu$ is the collision frequency and 
we have used the \citet{lenard58} collision operator, a natural modelling 
choice in this context because its eigenfunctions are Hermite polynomials. 

The free energy \exref{eq:W} in these terms is 
\beq
W = \frac{1+\alpha}{2}\la g_0^2\ra + \frac{1}{2}\sum_{m=1}^\infty \la g_m^2\ra 
\eeq
and satisfies
\beq
\frac{\rmd W}{\rmd t} = \frac{1+\alpha}{2}\,\eps - \nu\sum_{m=2}^\infty m\la g_m^2\ra.
\label{eq:Wbal}
\eeq

\subsection{FDR in Hermite space}
\label{sec:FDR_Hermite}

It is an obvious generalisation of the FDR to seek a relationship between 
the fluctuation level in the $m$-th Hermite moment, $\la|g_m|^2\ra$ 
(the ``Hermite spectrum''), and the injected power $\eps$.
This can be done in exactly the same manner as the kinetic FDR was 
derived in \secref{sec:FDR}. Hermite-transforming \eqref{eq:gko} gives   
\beq
\gmko = - \frac{i\chiko}{|k|}\frac{1+\alpha}{\alpha}\frac{(-\sgn k)^m}{\sqrt{2^m m!}}
\frac{Z^{(m)}(\omega/|k|)}{D_\alpha(\omega/|k|)},\quad m\ge1,
\label{eq:gm}
\eeq
where we have used 
\beq
Z^{(m)}(\zeta) \equiv \frac{\rmd^m Z}{\rmd\zeta^m} = 
(-1)^m\int\rmd v\,\frac{H_m(v) F_0(v)}{v-\zeta}
\label{eq:Zm}
\eeq
and $Z^{(m)}(\omega/k) = (\sgn k)^{m+1}Z^{(m)}(\omega/|k|)$. 
The mean square fluctuation level in the statistical steady state 
is then derived similarly to \eqref{eq:f}: 
\beq
\Cmk\equiv\la|\gmk|^2\ra = \frac{\eps_k}{2\pi|k|}\lt(\frac{1+\alpha}{\alpha}\rt)^2 
\frac{1}{2^m m!}\int_{-\infty}^{+\infty}\rmd\zeta\lt|\frac{Z^{(m)}(\zeta)}{D_\alpha(\zeta)}\rt|^2,
\quad m\ge1.
\label{eq:fm}
\eeq
This is the extension of the kinetic FDR, \eqref{eq:f}, to the fluctuations of the 
perturbed distribution function. The ``Hermite spectrum'' $\Cmk$ 
characterises the distribution of free energy in phase space. 

\subsection{Hermite spectrum}
\label{sec:spectrum}

It is interesting to derive the asymptotic form of this spectrum at $m\gg1$. 
Using in \eqref{eq:Zm} the asymptotic form of the Hermite polynomials at large $m$,
\beq
e^{-v^2/2} H_m(v) \approx 
\lt(\frac{2m}{e}\rt)^{m/2}
\sqrt{2}\cos\lt(v\sqrt{2m} - \pi m/2\rt),
\label{eq:Hm}
\eeq
and remembering that the $v$ integration is over the Landau contour 
(i.e., along the real line, circumnavigating the pole at $v=\zeta$ from below), 
we find
\beq
Z^{(m)}(\zeta) \approx i^{m+1}\sqrt{2\pi}\lt(\frac{2m}{e}\rt)^{m/2}e^{-\zeta^2/2 + i\zeta\sqrt{2m}},
\label{eq:Zm_approx}
\eeq 
provided $\zeta\ll\sqrt{2m}$ (this result is obtained by expressing the cosine 
in \eqref{eq:Hm} in terms of exponentials, completing the square in the
exponential function appearing in the integral \exref{eq:Zm} and moving 
the integration contour to $v = \pm i\sqrt{2m}$; the dominant contribution 
comes from the Landau pole). Finally, in \eqref{eq:fm}, 
\beq
\frac{|Z^{(m)}(\zeta)|^2}{2^m m!} \approx \sqrt{\frac{2\pi}{m}}\, e^{-\zeta^2},
\eeq
and so the Hermite spectrum has a universal scaling at $m\gg1$:
\beq
\Cmk \approx \lt[\frac{\eps_k}{\sqrt{2\pi}|k|}\lt(\frac{1+\alpha}{\alpha}\rt)^2 
\int_{-\infty}^{+\infty}\frac{\rmd\zeta\,e^{-\zeta^2}}{|D_\alpha(\zeta)|^2}\rt]\frac{1}{\sqrt{m}}
=\frac{\eps_k (1+\alpha)}{\sqrt{2 } |k|} \frac{1}{\sqrt{m}}. 
\label{eq:Cuniv}
\eeq 
The universal $1/\sqrt{m}$ scaling was derived in a different way by \citet{zocco11} 
(see \secref{sec:flux}; cf.~\citealt{watanabe04,hatch13}). 
The integral in \exref{eq:Cuniv} was evaluated using the Kramers--Kronig relations 
\citep{kramers27,kronig26} for the function
$h(\zeta) = 1/D_\alpha(\zeta) - \alpha$ 
(which is analytic in the upper half plane and decays at least as fast as $1/|\zeta|^2$ at large $\zeta$):
\beq
\int_{-\infty}^{+\infty}\frac{\rmd\zeta\,e^{-\zeta^2}}{|D_\alpha(\zeta)|^2} = 
-\sqrt{\pi}
\lt[\frac{1}{\pi}\,
{\cal P}\int_{-\infty}^{+\infty}\frac{\rmd\zeta\,\Imag\, h(\zeta)}{\zeta - \zeta'}\rt]_{\zeta'=0} 
= -\sqrt{\pi}\, \Real\, h(0) = \frac{\alpha^2}{1+\alpha}\sqrt{\pi}.
\eeq
Note that 
%this result is particularly easy to obtain 
%in the limit of zero real frequency and weak damping ($1+\alpha\ll1$, 
%the integral has a pole at $\zeta\ll1$ in \eqref{eq:Cuniv}; cf.~\secref{sec:small}),  
%as well as in the opposite limit of high frequency 
%($\alpha\gg1$, poles at $\zeta\gg1$; cf.~\secref{sec:large}). 
%In the latter limit, 
in the limit of high frequency ($\alpha\gg1$, \secref{sec:large}), 
the approximation \exref{eq:Zm_approx} requires $\omega_L/|k|\ll\sqrt{2m}$, or $\alpha\ll4m$,  
but there is also a meaningful intermediate range of $m$ for which 
$1\le m \ll \alpha/4$. In this range, we 
can approximate $Z(\zeta)\approx -1/\zeta$ and, since $\zeta\approx\pm\sqrt{\alpha/2}$, 
we have in \eqref{eq:fm}:
\beq
\frac{|Z^{(m)}(\zeta)|^2}{2^m m!} \approx \frac{2m!}{\alpha^{m+1}}
\quad\Rightarrow\quad
\Cmk \approx \frac{\eps_k}{\sqrt{\pi}|k|}\frac{m!}{\alpha^m}\,e^{\alpha/2}.
\eeq
This spectrum decays with $m$ up to $m\sim\alpha$, where it 
transitions into the universal spectrum \exref{eq:Cuniv}

\subsection{Free-energy flux, the effect of collisions and the FDR for the total free energy}
\label{sec:flux}

It could hardly have escaped a perceptive reader's 
notice that the total free energy in our system, with its $1/\sqrt{m}$ 
Hermite spectrum, is divergent. The regularisation in Hermite space 
(removal of fine velocity-space scales) is provided by collisions. 
If $\nu$ is infinitesimal, these are irrelevant at finite $m$, 
but eventually become important as $m\to\infty$. To take account 
of their effect and to understand the free-energy flow in 
Hermite space, we consider \eqref{eq:gmeq}, which it is convenient 
to Fourier transform in $z$ and rewrite in terms of 
$\tgmk \equiv (i\,\sgn k)^m\gmk$: 
\beq
\pd{\tgmk}{t} + \frac{|k|}{\sqrt{2}}\lt(\sqrt{m+1}\,\tg_{m+1,k} - \sqrt{m}\,\tg_{m-1,k}\rt) 
= - \nu m\tgmk. 
\label{eq:tg}
\eeq
The Hermite spectrum $\Cmk=\la|\gmk|^2\ra =\la|\tgmk|^2\ra$ therefore satisfies 
\beq
\pd{\Cmk}{t} + \Gamma_{m+1/2,k} - \Gamma_{m-1/2,k} = - 2\nu m \Cmk,
\label{eq:Cmk_exact}
\eeq
where $\Gamma_{m+1/2,k} = |k|\sqrt{2(m+1)}\Re\la\tg_{m+1,k}\tgmk^*\ra$ is the 
free-energy flux in Hermite space. If we make an assumption 
(verified in \secref{sec:cont})
that for $m\gg1$ the Hermite moments $\tgmk$ are continuous in $m$, i.e., 
$\tg_{m+1,k}\approx\tgmk$, then 
\beq
\Gamma_{m+1/2,k}\approx |k|\sqrt{2(m+1)}\,C_{m+1,k}
\label{eq:Gamma_cont}
\eeq
and \eqref{eq:Cmk_exact} turns into a closed evolution equation for the 
Hermite spectrum \citep{zocco11}: 
\beq
\pd{\Cmk}{t} + |k|\pd{}{m}\sqrt{2m}\,\Cmk = - 2\nu m \Cmk.
\label{eq:Cmk_approx}
\eeq
The universal $\Cmk\propto 1/\sqrt{m}$ spectrum derived in \secref{sec:spectrum} 
is now very obviously a constant-flux spectrum, reflecting steady 
pumping of free energy towards higher $m$'s (phase mixing). 
The full steady-state solution of 
\eqref{eq:Cmk_approx} including the collisional cutoff is 
\beq
\Cmk = \frac{A_k}{\sqrt{m}}\,\exp\lt(-\frac{2\sqrt{2}}{3}\frac{\nu}{|k|}m^{3/2}\rt),
\label{eq:Ccoll}
\eeq
where $A_k$ is an integration constant, which must be determined by matching 
this high-$m$ solution with the Hermite spectrum at low $m$. This we are now 
in a position to do: for $1\ll m\ll (\nu/|k|)^{-2/3}$, $\Cmk\approx A_k/\sqrt{m}$ 
and comparison with \eqref{eq:Cuniv} shows that the constant $A_k$ is the 
same as the constant $A_k(\alpha)$ in that equation. Thus, \eqref{eq:Ccoll} 
with $A_k$ given by \eqref{eq:Cuniv} provides a uniformly valid 
expression for the Hermite-space spectrum, including the collisional cutoff
(modulo the Hermite-space continuity assumption \exref{eq:Gamma_cont}, 
which we will justify in \secref{sec:cont}).  

As a check of consistency of our treatment, let us calculate 
the collisional dissipation rate of the free energy. 
This is the second term on the right-hand side of \eqref{eq:Wbal}. 
Since $\Cmk\propto1/\sqrt{m}$ before the collisional cutoff is reached, 
the sum over $m$ will be dominated by $m\sim(\nu/|k|)^{-2/3}$ 
and can be approximated by an integral: 
\beq
\nu\sum_{m,k} m\Cmk \approx \sum_k \nu\int_0^\infty\rmd m\,m\Cmk 
= \sum_k \frac{A_k|k|}{\sqrt{2}}. 
\eeq
On the other hand, in steady state, \eqref{eq:Wbal} implies 
\beq
\nu\sum_{m,k} m\Cmk = \frac{1+\alpha}{2}\,\eps. 
\label{eq:Wbal_stst}
\eeq
If energy injection is into a single $k$ mode, $\eps=\eps_k$,  
comparing these two expressions implies 
\beq
A_k = \frac{\eps_k (1+\alpha)}{\sqrt{2}|k|}, 
\label{eq:Ak}
\eeq
which, of course, is consistent with \eqref{eq:Cuniv}.

Finally, we use \eqref{eq:Ccoll} to calculate (approximately) 
the total steady-state amount of free energy across the phase space:
\beq
\frac{1}{2}\sum_{m=1}^\infty \Cmk 
= \frac{\Gamma(1/3)}{2\cdot3^{2/3}}\frac{1+\alpha}{\nu^{1/3}|k|^{2/3}}\,\eps_k
\label{eq:Wtot}
\eeq
(we have again approximated the sum with an integral, 
assumed energy injection into a single $k$ and used \eqref{eq:Ak}). 
\Eqref{eq:Wtot} can be thought of as the FDR for the total free energy.  
The fact that this diverges as $\nu\to 0$ 
underscores the principle that the ``true'' dissipation (in the sense of free 
energy being thermalised) is always collisional---a consequence of Boltzmann's 
$H$ theorem.  

\subsection{Continuity in Hermite space}
\label{sec:cont}

In this section, we make a somewhat lengthy formal digression to 
justify the assumption of continuity of Hermite moments in $m$ at 
large $m$, which we need for the approximation \exref{eq:Gamma_cont}. 
The formalism required for this will have some interesting features 
which are useful in framing one's thinking about energy flows in 
Hermite space, but a reader impatient with such exercises may 
skip to \secref{sec:LF}. 

Returning to \eqref{eq:tg} and considering $1\ll m\ll (\nu/|k|)^{-2}$, 
we find that to lowest approximation, the $\sqrt{m}$ terms are dominant 
and must balance, giving $\tg_{m+1,k}\approx\tg_{m-1,k}$. 
This is consistent with continuity in $m$, viz., $\tg_{m+1,k}\approx\tg_{m,k}$, 
but there is also a solution allowing the consecutive Hermite 
moments to alternate sign: $\tg_{m+1,k}\approx-\tgmk$. 
Thus, there are, formally speaking, two solutions: one for which 
$\tgmk$ is continuous and one for which $(-1)^m\tgmk$ is. To take into account 
both of them, we introduce the following decomposition 
\citep{schekochihin14}: 
\beq
\tgmk = \tgmk^+ + (-1)^m\tgmk^-,
\label{eq:decomp}
\eeq
where the ``$+$'' (``continuous'') and the ``$-$'' (``alternating'') modes are
\beq
\tgmk^+ = \frac{\tgmk + \tg_{m+1,k}}{2}, \quad
\tgmk^- = (-1)^m\frac{\tgmk - \tg_{m+1,k}}{2}.
\label{eq:gpm}
\eeq
The Hermite spectrum and the flux of the free energy can be expressed 
in terms of the spectra of these modes as follows: 
\begin{align}
&\Cmk \equiv \la|\tgmk|^2\ra = \Cmk^+ + \Cmk^-,\\
&\Gamma_{m+1/2,k} \equiv |k|\sqrt{2(m+1)}\Re\la\tg_{m+1,k}\tgmk^*\ra 
\approx |k|\sqrt{2m}\lt(\Cmk^+ - \Cmk^-\rt),
\label{eq:Gamma_pm}
\end{align}
where $\Cmk^\pm\equiv \la|\tgmk^\pm|^2\ra$ and the last expression 
in \eqref{eq:Gamma_pm} is an approximation valid for $m\gg1$. 

The functions $\tgmk^\pm$ can both be safely treated as continuous in $m$ for $m\gg1$. 
Treating them so in \eqref{eq:tg} and working to lowest order in $1/m$, 
we find that they satisfy the following {\em decoupled} evolution equations: 
\beq
\pd{\tgmk^\pm}{t} \pm \sqrt{2}|k| m^{1/4}\pd{}{m} m^{1/4}\tgmk^\pm = - \nu m\tgmk^\pm,
\eeq 
or, for their spectra, 
\beq
\pd{\Cmk^\pm}{t} \pm |k|\pd{}{m}\sqrt{2m}\,\Cmk^\pm = - 2\nu m \Cmk^\pm.
\eeq
Manifestly, the ``$+$'' mode propagates from lower to higher $m$ and 
the ``$-$'' mode from higher to lower $m$---they are the ``phase-mixing'' and 
the ``un-phase-mixing'' collisionless solutions, respectively.\footnote{The existence 
of un-phase-mixing solutions has been known for a long time: e.g., 
\citet{hammett93} treated them as forward and backward propagating 
waves in a mechanical analogy of \eqref{eq:tg} with a row of masses 
connected by springs. The un-phase-mixing solutions are also 
what allows the phenomenon of plasma echo \citep{gould67}, including 
in stochastic nonlinear systems \citep{schekochihin14}.}

Taking the collisional term into account and noting that energy is injected 
into the system at low, rather than high, $m$, the solution satisfying 
the boundary condition $\tgmk\to 0$ as $m\to\infty$ has $\tgmk^-=0$
and so $\tgmk=\tgmk^+$. Thus, $\tgmk$ is continuous in $m$. 
With $\Cmk^-=0$, \eqref{eq:Gamma_pm} is the same as our earlier 
approximation \exref{eq:Gamma_cont} (to lowest order in the $m\gg1$ expansion). 

As $\tgmk^+$ and $\tgmk^-$ are decoupled at large $m$, if we start with a 
$\tgmk^-=0$ solution, no $\tgmk^-$ will be produced. 
However, both the decoupling property and the interpretation of 
$\tgmk^\pm$ as the phase-mixing and un-phase-mixing modes
are only valid to lowest order in $1/m$. It is useful 
to know how well this approximation holds. 

\begin{figure}
\begin{center}
\includegraphics[width=10cm]{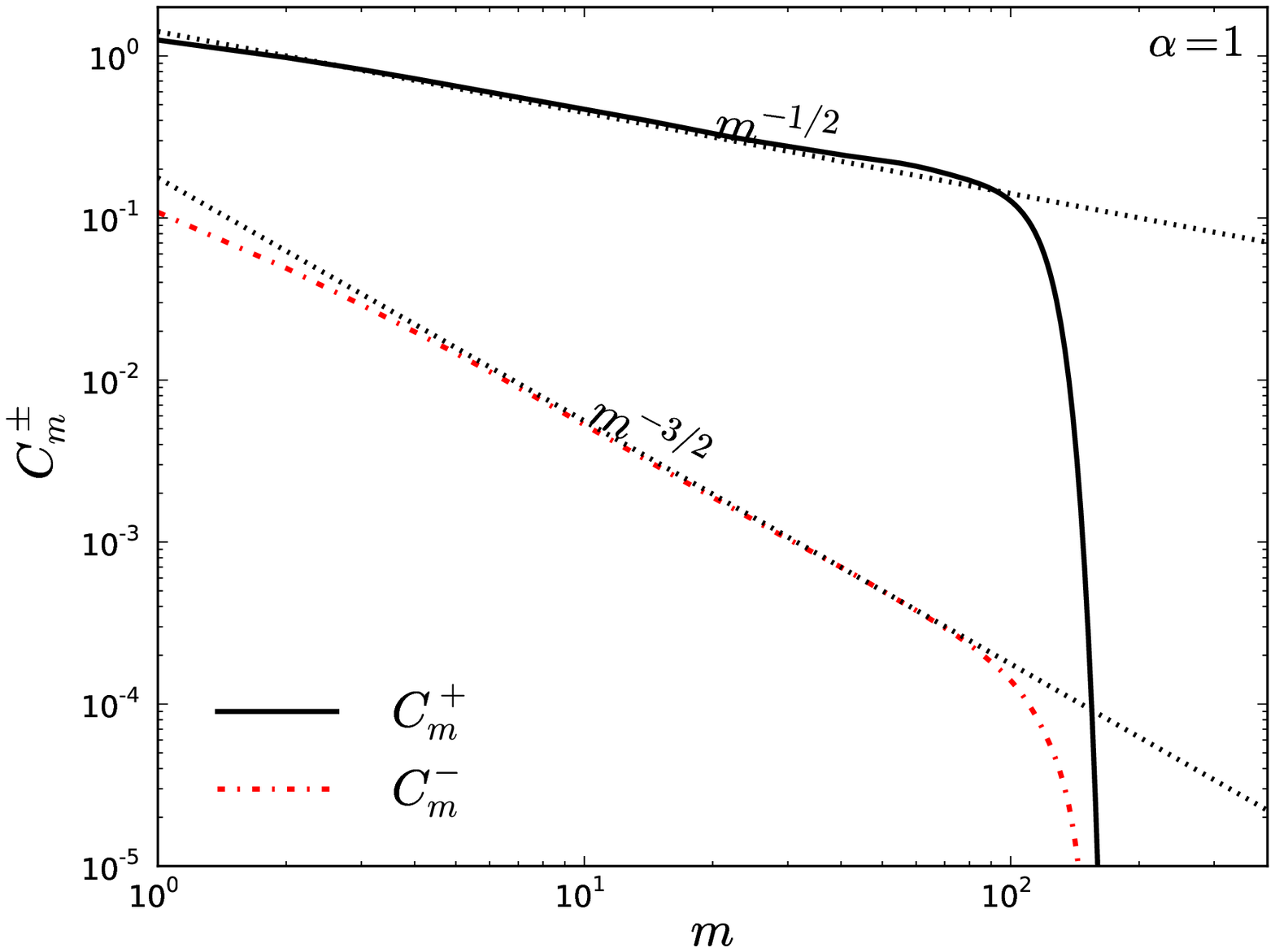}
\caption{The free-energy spectra $C_m^\pm$ obtained via direct numerical solution 
of \eqsdash{eq:g0}{eq:gmeq} with $\alpha = 1.0$ followed by decomposing the solution 
according to \eqref{eq:gpm}. In the code, rather than using the Lenard--Bernstein 
collision operator (as per \eqref{eq:gmeq}), hypercollisional regularisation \citep{loureiro13},
$-\nu m^6\gmk$, was used to maximise the utility of the velocity-space resolution, 
hence the very sharp cut off. 
The dotted lines show the collisionless approximation: \eqref{eq:Cuniv} 
for $\Cmk^+$ (the phase-mixing ``$+$'' mode predominates, so 
$\Cmk\approx\Cmk^+$) and \eqref{eq:Cminus} for $\Cmk^-$.} 
\label{fig:Cpm}
\end{center}
\end{figure}

Let us use \eqref{eq:gm} to calculate (in the collisionless limit) 
\beq
R_{m+1} \equiv \frac{\tg_{m+1,k\omega}}{\tgmko} = 
i\,\sgn k\,\frac{g_{m+1,k\omega}}{\gmko} 
= -\frac{i}{\sqrt{2(m+1)}}\frac{Z^{(m+1)}(\zeta)}{Z^{(m)}(\zeta)}.
\label{eq:Rm_exact}
\eeq
Taking $m\gg1, \zeta^2/4$ and using \eqref{eq:Zm_approx}, we find\footnote{The same 
lowest-order expression can be found by Fourier-transforming 
\eqref{eq:tg} in time, ignoring collisions, 
writing $R_{m+1} = R_m^{-1}\sqrt{m/(m+1)} + i\zeta\sqrt{2/(m+1)}$, 
approximating $R_m\approx R_{m+1}$, solving the resulting quadratic equation for $R_{m+1}$,  
expanding in powers of $1/\sqrt{m}$ and choosing the solution for which $R_{m+1}=1$ 
to lowest order. This last step is the main difference between the two methods: 
if we work with \eqref{eq:tg} in the manner just described, we have to make an explicit 
choice between the continuous and alternating solutions ($R_{m+1}=1$ and $R_{m+1}=-1$); 
on the other hand, \eqref{eq:gm} already contains the choice of the former (which is ultimately 
traceable to Landau's prescription guaranteeing damping rather than growth of the perturbations).}
\beq
R_{m+1} = 1 + \frac{i\zeta}{\sqrt{2m}} - \frac{1}{4m} + O\lt(\frac{1}{m^{3/2}}\rt).
\label{eq:Rm_approx}
\eeq
Therefore, to lowest order in $1/\sqrt{m}$, 
\beq
\tgmko^- = (-1)^m\tgmko\,\frac{1-R_{m+1}}{2} 
\approx(-1)^{m+1}\,\frac{i\zeta}{2\sqrt{2m}}\,\tgmko. 
\eeq
Following the same steps as those that led to \eqref{eq:Cuniv}\footnote{The integral is 
again calculated via Kramers--Kronig relations, this time for the function 
$h(\zeta) = \zeta^2/D_\alpha(\zeta) - \alpha \zeta^2 - \alpha^2/2$, 
so $\int_{-\infty}^{+\infty}\rmd\zeta\,\zeta^2 e^{-\zeta^2}\!/|D_\alpha(\zeta)|^2
= \alpha^2\sqrt{\pi}/2$.}, 
we get
\beq
\Cmk^- \approx \lt[\frac{\eps_k}{8\sqrt{2\pi}|k|}\lt(\frac{1+\alpha}{\alpha}\rt)^2 
\int_{-\infty}^{+\infty}\frac{\rmd\zeta\,\zeta^2
e^{-\zeta^2}}{|D_\alpha(\zeta)|^2}\rt]\frac{1}{m^{3/2}} 
 = \frac{\eps_k (1+\alpha)^2}{16 \sqrt{2} |k|} \frac{1}{m^{3/2}}, 
\label{eq:Cminus}
\eeq
so both the energy ($\sim 1$, while the total is $\sim\nu^{-1/3}$; see \eqref{eq:Wtot})
and the dissipation ($\sim\nu\sum_m m\Cmk^-\sim \nu^{2/3}$) associated with the 
``$-$'' modes are small.  

The steady-state spectra $\Cmk^\pm$ obtained via direct numerical solution 
of \eqsand{eq:g}{eq:phi} are shown in \figref{fig:Cpm}, where they are also 
compared with the analytical expressions \exref{eq:Cuniv} and \exref{eq:Cminus}.  

Note that we could have, without further ado, simply taken 
\eqref{eq:Rm_approx} to be the proof of continuity in Hermite space. We have chosen 
to argue this point via the decomposition \exref{eq:decomp} because it provided 
us with a more intuitive understanding of the connection between this continuity 
and the direction of the free-energy flow (phase mixing rather than un-phase mixing). 

\subsection{The simplest Landau-fluid closure}
\label{sec:LF}

Simplistically described, the idea of Landau-fluid closures 
is to truncate the Hermite hierarchy of \eqsdash{eq:g0}{eq:gmeq} 
at some finite $m$ and to replace in the last retained equation
\beq
g_{m+1,k}(t) = -(i\,\sgn k) R_{m+1}\gmk(t),
\label{eq:trunc}
\eeq
where $R_{m+1}$, which in general depends on the complex frequency $\zeta$ (\eqref{eq:Rm_exact}),
is approximated by some suitable frequency-independent expression leading 
to the correct recovery of the linear physics from the truncated system.   
A considerable level of sophistication has been achieved in making these choices 
and we are not proposing to improve on the existing literature 
\citep{hammett90,hammett92,hedrick92,dorland93,snyder97,passot04,goswami05,passot07}. 
It is, however, useful, in the context of the result of \secref{sec:small} that 
the ``fluid'' version of FDR is recovered in the limit of low frequency and weak 
damping, to show how the same conclusion can be arrived at via what is probably 
the simplest possible Landau-fluid closure. 

In the limit $\zeta\to0$, the ratio $R_{m+1}$, given by \eqref{eq:Rm_exact}, becomes 
independent of $\zeta$ and so a closure in the form \exref{eq:trunc} 
becomes a rigorous approximation. It is not hard to show that 
\beq
Z^{(m)}(0) = \frac{i^{m+1}\sqrt{\pi}\, m!}{\Gamma(m/2 + 1)}.
\eeq
Therefore, for $\zeta\ll1$ and $m\ge1$,\footnote{The same result can be obtained 
by inferring $R_{m+1} \approx R_m^{-1}\sqrt{m/(m+1)}$ from \eqref{eq:tg} 
(provided $m\ll1/\zeta^2$), then iterating this up to some Hermite number $M$
such that $1\ll M\ll1/\zeta^2$, and approximating $R_M\approx1$ (\eqref{eq:Rm_approx}).
The condition $m, M \ll 1/\zeta^2$ is necessary so that the $\zeta$ terms in 
$R_{m+1}$ are not just small compared to unity but also compared to the next-order 
$1/m$ terms (see \eqref{eq:Rm_approx}).} 
\beq
R_{m+1} = \frac{m}{\sqrt{2(m+1)}}\frac{\Gamma(m/2)}{\Gamma((m+1)/2)}. 
\eeq
If we wish to truncate at $m=1$, then $R_2=\sqrt{\pi}/2$, and so in \eqref{eq:g1}, 
\beq
g_{2,k}=-i\,\sgn k\,\frac{\sqrt{\pi}}{2}\,g_{1,k}.
\eeq
On the basis of \eqref{eq:g0}, we must order $g_{1,k}\sim O(\zeta) g_{0,k}$. 
Thefore, $\dd g_{1,k}/\dd t \sim O(\zeta^2) g_{0,k}$ must be neglected 
in \eqref{eq:g1}, from which we then learn that 
\beq
g_{1,k}\approx -i\,\sgn k\,\sqrt{\frac{2}{\pi}}\lt(1+\alpha\rt) g_{0,k}.
\eeq
Finally, substituting this into \eqref{eq:g0}, we get
\beq
\pd{g_{0,k}}{t} + \frac{1+\alpha}{\sqrt{\pi}}|k| g_{0,k} = \chi_k.
\label{eq:LF_Langevin}
\eeq
This is a Langevin equation \exref{eq:Langevin} with a damping rate that 
is precisely the Landau damping rate $\gamma_L$ in the limit $1+\alpha\ll1$ 
(and so $\zeta\ll1$), given by \eqref{eq:omega_small}. 
In this limit, $\phi = -g_0$ (\eqref{eq:phi_g0}, $\alpha\approx-1$) 
and we recover the standard ``fluid'' FDR (\eqref{eq:FDR_small}). 
As we discussed in \secref{sec:kin}, a useful application of this regime 
is to compressive fluctuations in high-beta plasmas: in this case 
$1+\alpha\approx 1/\beta_i\ll1$ and the damping is the \citet{barnes66}
damping, well known in space and astrophysical contexts \citep{foote79,lithwick01,tome}.

\section{Conclusion}
\label{sec:disc}

We have provided what in our view is a reasonably complete treatment 
of the simplest generalisation of the Langevin problem to plasma kinetic 
systems.\footnote{While we have focused on the simplest Langevin problem, in which the 
source term is a white noise, there is an obvious route towards generalising 
this by considering source terms with more coherent time dependence 
(longer correlation times, prescribed frequency spectra; cf.\ \citealt{plunk13}). 
One such calculation was recently undertaken by \citet{plunk14}, who 
considered a coherent oscillating source and found
that when the frequency of the source is large, the amount of energy that 
can be absorbed by the kinetic system is exponentially small
(which makes sense). Another straightforward generalisation (or variation) of our model
(as treated in the main text of this article) 
is energy injection into momentum, rather than density fluctuations---which can 
be interpreted as forcing by an externally imposed random electric field.
Whereas some of the more literal parallels with the Langevin problem 
are lost in this case, the results are fundamentally the same (\apref{ap:g1force}).} 
Let us itemise the main results and conclusions.\\ 

\begin{itemize}

\item \Eqref{eq:f} is the FDR for the kinetic system (\eqsand{eq:g}{eq:phi}), 
expressing the relationship between the fluctuation level $\la|\phi_k|^2\ra$ 
and the injected power. This can be expressed in terms of an ``effective'' 
damping rate $\geff$ in a way that resembles the standard ``fluid'' version 
of the FDR (\eqref{eq:gamma}), but $\geff$ is not in general equal to the 
Landau damping rate $\gamma_L$. We stress that this result is not a statement 
of any kind of surprising ``modification'' of Landau damping in a system with 
a random source, but rather a clarification of what the linear response 
in the statistical steady state of such a system actually is. The system, in general, 
is not mathematically equivalent to the Langevin equation \exref{eq:Langevin}
and so the FDR for it need not have the same form. 

\item In the limit of zero real frequency and weak Landau damping, the 
effective and the Landau damping rates do coincide (\eqref{eq:FDR_small}). 
Another way to view this result is by noting that 
this is a regime in which the simplest possible Landau-fluid closure 
becomes a rigorous approximation and the evolution equation for the 
electrostatic potential can be written as a Langevin equation with 
the Landau damping rate $\gamma_L$ (\eqref{eq:LF_Langevin}). Note, however, 
that this direct reduction to the simplest Langevin equation \exref{eq:Langevin} 
is not a universal property: it breaks down with a different choice 
of forcing (\apref{ap:g1force}). 

\item Another limit in which the FDR for the kinetic system can be 
interpreted in ``fluid'' (in fact, mechanical) terms is one of high real 
frequency and exponentially Landau small damping, although the correct 
analogy is not the Langevin equation but a forced-damped 
oscillator (\secref{sec:large}; this analogy, however, ceases to hold 
in such a simple form for a different choice of forcing, as shown in \apref{ap:g1force}).

\item The damping of the perturbations of $\phi$ (which are linearly 
proportional to the density perturbations) occurs via phase mixing, which 
transfers the free energy originally injected into $\phi$ away from it 
and into higher moments of the perturbed distribution function. This 
process can be described as a free-energy flow in Hermite space. 
The generalisation of the FDR to higher-order Hermite moments 
takes the form of an expression for the Hermite spectrum $\Cmk$ 
(\eqref{eq:fm}), which at high Hermite numbers $m\gg1$ has a universal 
scaling $\Cmk\propto 1/\sqrt{m}$ (\eqref{eq:Cuniv}). This scaling 
corresponds to a constant free-energy flux from low to high $m$ 
(\eqref{eq:Gamma_cont}). Analysis of the solutions of the kinetic equation 
making use of a formal decomposition of these solutions into phase-mixing 
and un-phase-mixing modes underscores the predominance of the former 
(\secref{sec:cont}). 

\item A solution for the Hermite spectrum including the collisional 
cutoff is derived (\eqref{eq:Ccoll}). The FDR for the total free 
energy stored in the phase space (\eqref{eq:Wtot}) shows that it 
diverges $\propto\nu^{-1/3}$ in the limit of vanishing collisionality 
$\nu$, a result that underscores the fact that ultimately all 
dissipation (i.e., all entropy production in the system) is collisional.\\ 

\end{itemize}

In the process of deriving all these results, we have made an effort 
to explain the simple connections between the Landau formalism 
(solutions of the kinetic equation expressed via the plasma dispersion function)
and the Hermite-space one. 
This material and, indeed, most of the results described above, 
perhaps belong to elementary textbooks, but we are not aware of any where 
they are adequately explained---although implicitly they underlie 
the thinking behind both Landau-fluid closures 
\citep{hammett90,hammett92,hedrick92,dorland93,snyder97,passot04,goswami05,passot07}
and Hermite-space treatments for plasma kinetics
\citep{armstrong67,grant67a,hammett93,parker95,ng99,watanabe04,zocco11,loureiro13,hatch13,plunk14}.

Besides (we hope) providing a degree of clarity on an old 
topic in the linear theory of collisionless plasmas, our findings lay the groundwork for 
a study of the much more complicated nonlinear problem of the role of Landau 
damping and phase mixing in turbulent collisionless plasma systems 
\citep{schekochihin14,kanekar14}.

%%%%%%%%%%%%%%%%%%%%%%%%%%%%%%%%%%%%%%%%%%%%%%%%%%%%%%%%%%%%%%%%%%%%%%%%%%%%%%%%%%%%%%%%%%%%%%%%%%%%%%%
\begin{acknowledgments}

The authors are grateful for fruitful discussions with 
T.\ Antonsen, M.\ Barnes, P.\ Dellar, J.\ Drake, A.\ Hassam, J.\ Parker, G.\ Plunk, 
J.\ TenBarge, A.\ Zocco, and especially G.\ Hammett. 
We would also like to thank an anonymous referee for a very thorough 
report, which led to significant improvements in our presentation. This material is based
upon work supported by the U.S. Department of Energy, Office of Science, Office of Fusion
Energy Science, under Award Numbers DEFG0293ER54197 and DEFC0208ER54964. The work of AAS
was carried out, in part, within the framework of the EUROfusion Consortium and has
received funding from the European Union’s Horizon 2020 research and innovation programme
under Grant Agreement No 633053. The views and opinions expressed herein do not
necessarily reflect those of the European Commission.
NFL was also partially supported by Funda\c{c}\~{a}o para a
Ci\^{e}ncia e Tecnologia through grants Pest-OE/SADG/LA0010/2011, IF/00530/2013 and
PTDC/FIS/118187/2010. 
The numerical simulations were carried out on the Dirac machine of the National 
Energy Research Scientific Computing Center, which is supported by the Office of 
Science of the U.S.\ Department of Energy under Contract No.\ DE-AC02-05CH11231.
    
\end{acknowledgments}

\appendix

\section{Momentum forcing}
\label{ap:g1force}

The source term in \eqref{eq:g}, providing direct forcing of density perturbations, 
was a choice of convenience: it allowed us to compare directly the FDR
for the potential field $\phi$ in a kinetic system with the FDR for 
the Langevin equation \exref{eq:Langevin}. If, instead, one strives for 
a form of energy injection with a more transparent physical interpretation, 
it is natural to imagine it 
coming from a fluctuating electric field. This changes \eqref{eq:g} 
to the following:
\begin{align}
&\pd{g}{t} + v\,\pd{g}{z} 
+ v F_0 \pd{\phi}{z} = 
\chi_1(t) v F_0 + C[g] \;\!\!, \label{eq:g1force} \\
& \la \chi_1(t) \chi_1(t') \ra  = \eps \delta (t-t'), \nonumber
\end{align}
where $\chi_1(t)$ is the fluctuating parallel electric field, which we
model (again, for analytical convenience) as a Gaussian white noise. 

The new forcing injects fluctuations of momentum, rather than density. 
Indeed, in terms of Hermite moments, instead of \eqsand{eq:g0}{eq:g1}, 
we now have 
\begin{align} 
\label{eq:g0mom}
&\pd{g_0}{t} + \pd{}{z}\frac{g_1}{\sqrt{2}}  = 0,\\
\label{eq:g1mom}
&\pd{g_1}{t} + \pd{}{z}\lt(g_2 + \frac{1+\alpha}{\sqrt{2}}\,g_0\rt)  = \frac{\chi_1}{\sqrt{2}},
\end{align}
and \eqref{eq:gmeq} is unchanged. 
The field that is directly forced is $g_1 = \sqrt{2}\int \rmd v\,v g(v)$, which is 
proportional to the mean velocity associated with the perturbed distribution $g$. 
The new free-energy equation, an analogue of \eqsand{eq:Wbalance}{eq:Wbal}, is 
\beq
\frac{\rmd W}{\rmd t} = \frac{\eps}{4} + \int\rmd v\,\frac{\la g C[g]\ra}{F_0}
= \frac{\eps}{4} - \nu \sum_{m=2}^\infty m\la g_m^2\ra.
\label{eq:Wbal1}
\eeq
This immediately gives us the universal Hermite spectrum and the FDR for the total 
free energy: we repeat the calculation in \secref{sec:flux} (which is unchanged 
because nothing has changed at high $m$'s) using the steady-state 
version of \eqref{eq:Wbal1} instead of \eqref{eq:Wbal_stst} to get 
\beq
A_k = \frac{\eps_k}{2\sqrt{2}|k|} 
\label{eq:Ak1}
\eeq
in the expression \exref{eq:Ccoll} for the Hermite spectrum. 
Therefore,   
\beq
\frac{1}{2}\sum_{m=1}^\infty \Cmk 
= \frac{\Gamma(1/3)}{4\cdot3^{2/3}}\frac{1}{\nu^{1/3}|k|^{2/3}}\,\eps_k
\label{eq:Wtot1}
\eeq
replaces \eqref{eq:Wtot} as the FDR for the total free energy. 
The only differences are in numerical prefactors and the $\alpha$ dependence, 
which has now disappeared. This is because in our previous 
forcing model, the source term injected
energy into $g_0$ (density fluctuations), which got scaled by the factor of $1+\alpha$ 
when passed on to $g_1$ (see \eqref{eq:g1}), 
whereas in the case we are considering now, the energy is injected directly into $g_1$, 
which is then phase mixed to higher $m$'s, without ever encountering any $\alpha$ dependence.  

\begin{figure}
\begin{center}
\includegraphics[width=10cm]{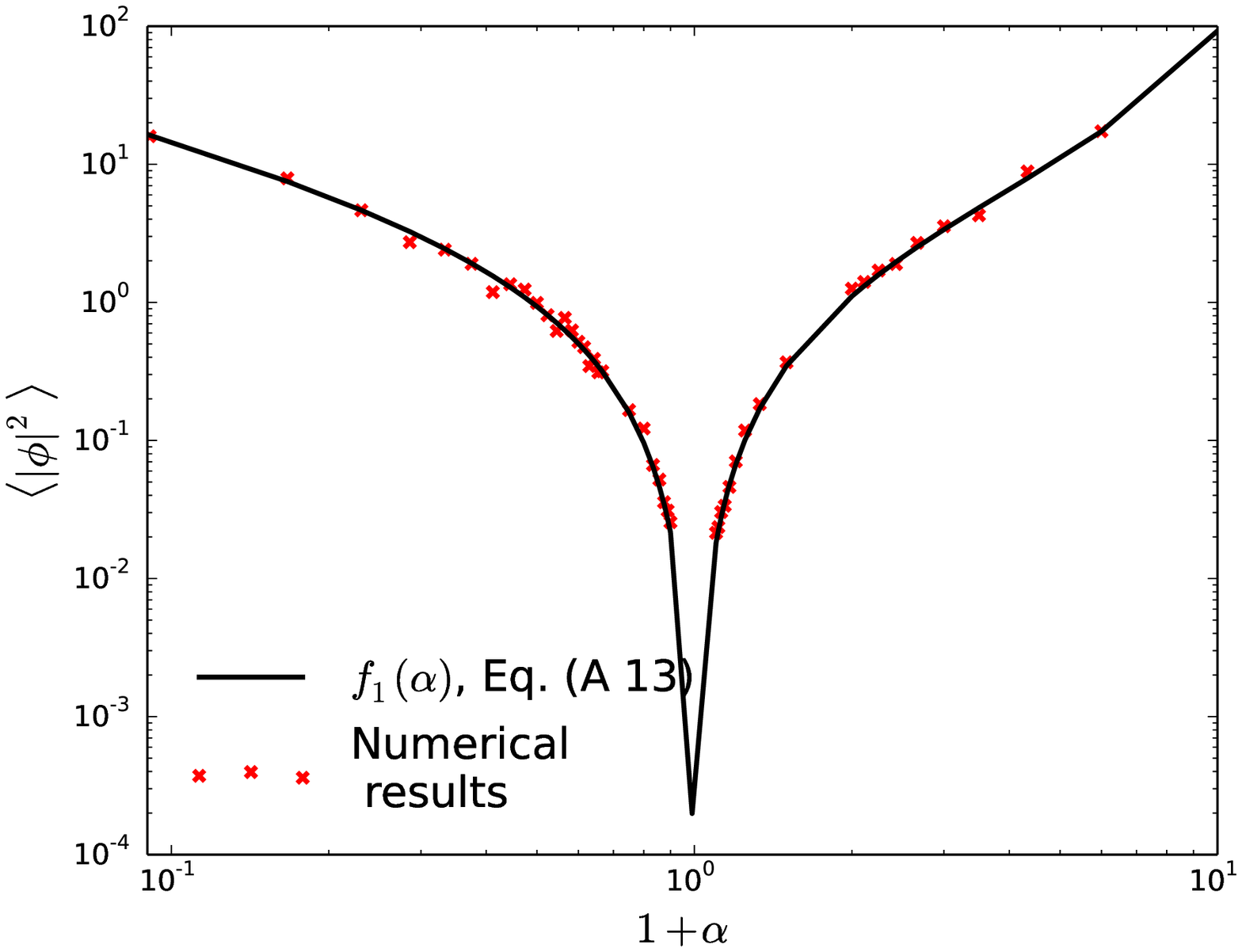}
\caption{Normalised steady-state amplitude $2\pi|k|\la|\phi_k|^2\ra/\eps_k=f_1(\alpha)$ 
vs.\ $1+\alpha$ for the case of momentum forcing: 
the solid line is the analytical prediction $f_1(\alpha)$ (\eqref{eq:f1}), 
the crosses are computed from 
the long-time limit of $\la|\phi_k|^2\ra$ obtained via direct numerical 
solution of \eqref{eq:g1force}.}
\label{fig:f1}
\end{center}
\end{figure}

Let us also give here the results one obtains in the collisionless 
limit by backtracking to \eqref{eq:g1force} and solving for $g$ explicitly, 
as we did in \secsand{sec:FDR}{sec:FDR_Hermite}:
\beq
\gko = - \lt(\phiko + \frac{i\chikoone}{k}\rt)\frac{v F_0}{v - \omega/k}.
\label{eq:gko1}
\eeq
This gives
\begin{align}
\label{eq:phiko1}
\phiko &= - \frac{i\chikoone}{k}\frac{1 + \zeta Z(\zeta)}{D_\alpha(\zeta)},\\
\gmko &= - \frac{i\chikoone}{k}\frac{1}{\alpha}\frac{(-\sgn k)^m}{\sqrt{2^m m!}}
\frac{\zeta Z^{(m)}(\zeta)}{D_\alpha(\zeta)},\quad m\ge1,
\label{eq:gm1}
\end{align}
where $\zeta = \omega/|k|$ as usual. From the last formula, 
proceeding in the same manner as we did to get \eqref{eq:Cuniv}, we recover again
the Hermite spectrum: 
\begin{align}
\Cmk &= \frac{\eps_k}{2\pi|k|}\frac{1}{\alpha^2} 
\frac{1}{2^m m!}\int_{-\infty}^{+\infty}\rmd\zeta\lt|\frac{\zeta Z^{(m)}(\zeta)}{D_\alpha(\zeta)}\rt|^2\\
&\approx \lt[\frac{\eps_k}{\sqrt{2\pi}|k|}\frac{1}{\alpha^2}
\int_{-\infty}^{+\infty}\frac{\rmd\zeta\,\zeta^2 e^{-\zeta^2}}{|D_\alpha(\zeta)|^2}\rt]\frac{1}{\sqrt{m}}
=\frac{\eps_k}{2\sqrt{2 } |k|} \frac{1}{\sqrt{m}}. 
\label{eq:Cuniv1}
\end{align}
The latter expression was obtained in the limit of $m\gg1$ (see \secref{sec:spectrum}) 
and is the same result as \eqref{eq:Ak1}. The integral is already familiar from \eqref{eq:Cminus}. 
For completeness, the ``$-$''-mode spectrum \exref{eq:Cminus} becomes 
\beq
\Cmk^- \approx \lt[\frac{\eps_k}{8\sqrt{2\pi}|k|}\frac{1}{\alpha^2}
\int_{-\infty}^{+\infty}\frac{\rmd\zeta\,\zeta^4
e^{-\zeta^2}}{|D_\alpha(\zeta)|^2}\rt]\frac{1}{m^{3/2}} 
 = \frac{\eps_k (3+\alpha)}{32 \sqrt{2} |k|} \frac{1}{m^{3/2}}. 
\label{eq:Cminus1}
\eeq
The integral was done by Kramers--Kroning relations
for the function $h(\zeta) = \zeta^4/D_\alpha(\zeta) - \alpha\zeta^4 - \alpha^2\zeta^2/2 
- \alpha^2(3+\alpha)/4$. While again numerical prefactors and $\alpha$ dependence 
are different, none of the substative arguments in \secref{sec:cont} are affected. 

Finally, from \eqref{eq:phiko1}, 
proceeding in the same manner as in \secref{sec:FDR}, we obtain the FDR 
relation for the mean square fluctuation amplitude of the potential:
\beq
\la |\phi_k|^2 \ra = \frac{\eps_k }{2 \pi |k|} f_1(\alpha), \quad
f_1(\alpha) = \int_{-\infty}^{+\infty}\rmd\zeta
\lt|\frac{1 + \zeta Z(\zeta)}{D_\alpha(\zeta)}\rt|^2,
\label{eq:f1}
\eeq
which is the new version of \eqref{eq:f}. 
The function $f_1(\alpha)$ is plotted in \figref{fig:f1}, along with
the results of the direct numerical solution of \eqref{eq:g1force}.
While formally it is a different function than $f(\alpha)$, 
it exhibits very similar behaviour (cf.\ \figref{fig:f}).  
Its asymptotics are (cf.\ \secsand{sec:small}{sec:large})
\begin{align}
\label{eq:FDR_small1}
\alpha\to-1: & \quad
f_1(\alpha) \approx \frac{|k|}{\gamma_L}
\quad\Rightarrow\quad
\la|\phi_k|^2\ra\approx \frac{\eps_k}{2\pi \gamma_L},\\
\alpha\to\infty: & \quad
f_1(\alpha) \approx \frac{\pi\alpha |k|}{4 \gamma_L}
\quad\Rightarrow\quad
\la|\phi_k|^2\ra\approx \frac{\alpha \eps_k}{8 \gamma_L}.
\label{eq:FDR_large1}
\end{align}
Whereas in both limits there is still an inverse relationship between 
the mean square fluctuation amplitude and the Landau damping rate 
$\gamma_L$, the numerical coefficients are not easily interpretable 
in terms of any simple ``fluid'' Langevin models for $\phi$---not a
surprising outcome as, already examining \eqsand{eq:g0mom}{eq:g1mom}, 
we might have observed that they do not map on any obvious 
Langevin-like equation for $\phi=\alpha g_0$.
%\footnote{The closest 
%to a Langevin-like form one gets is by rewriting these equations 
%in terms of two ``modes'' $u_\pm = g_1 \pm \sqrt{1+\alpha}\, g_0$, which satisfy 
%$\dd_t u_\pm \pm \omega_0 \dd_z u_\pm + \dd_z g_2 = \chi_1/\sqrt{2}$, 
%where $\omega_0=\sqrt{(1+\alpha)/2}$. This system has a source, a real frequency 
%$\omega_0$, and a damping that enters via the phase-mixing term $\dd_z g_2$.
%This damping is not, however, representable as $\gamma_L u_\pm$.} 
The elementary Landau-fluid closure that in \secref{sec:LF} neatly 
mapped the $\alpha\to-1$ limit onto a ``fluid'' Langevin equation, 
when reworked for the case of the momentum forcing, gives
\beq
\pd{\phi_k}{t} + \gamma_L \phi_k = \frac{\sgn k}{\sqrt{\pi}}\chi_{1,k}.
\label{eq:LF1}
\eeq
Thus, a Langevin equation still, but with an order-unity adjusted noise term.  

\bibliographystyle{jpp}
\bibliography{ksdl_JPP}

\end{document}